\documentclass[a4paper]{article}

\usepackage{natbib}

\usepackage{amsmath,amsfonts,latexsym,comment}
\usepackage[margin = 1in]{geometry}
\usepackage{float}
\usepackage{caption}
\usepackage{subcaption}
\usepackage{multirow}
\usepackage{diagbox}
\usepackage{ulem}
\usepackage{dsfont}
\usepackage{hyperref}
\usepackage{graphicx}
\usepackage{lastpage}
\usepackage{xcolor}

\hypersetup{
	colorlinks = true,
	citecolor = blue,
	linkcolor = blue
	}

\renewcommand{\phi}{\varphi}

\newcommand{\R}{\mathbb{R}}
\renewcommand{\epsilon}{\varepsilon}

\def\norm#1{\left\| #1 \right\|}

\def\acc#1{\left\{ #1 \right\}}
\def\pa#1{\left( #1 \right)}

\def\Ybf{\mathbf{Y}}
\def\Ybftild{\widetilde{\mathbf{Y}}}
\def\Xbf{\mathbf{X}}
\def\Xbftild{\widetilde{\mathbf{X}}}
\def\wbeta{\widehat\beta}
\def\Wtild{\widetilde W}
\def\bkR{\mathbb{R}}
\def\diag{\mbox{\rm diag}}
\def\mgm3{$\mu$g$\cdot$m$^{-3}$}
\newcommand{\celsius}{$^\circ$}

\begin{document}

\title{Geographically Weighted Regression for Air Quality Low-Cost Sensor Calibration}

\author{Jean-Michel Poggi, Bruno Portier and Emma Thulliez}
\maketitle

\begin{abstract}
This article focuses on the use of Geographically Weighted Regression (GWR) method to correct air quality low-cost sensors measurements. Those sensors are of major interest in the current era of high-resolution air quality monitoring at urban scale, but require calibration using reference analyzers. The results for NO2 are provided along with comments on the estimated GWR model and the spatial content of the estimated coefficients. The study has been carried out using the publicly available SensEURCity dataset in Antwerp, which is especially relevant since it includes 9 reference stations and 34 low-cost sensors collocated and deployed within the city.
\end{abstract}


\baselineskip=14.5pt    

\section{INTRODUCTION}

Low-cost sensors are a new tool for improving air quality maps, 
which are of major interest in the current era of high-resolution 
air quality monitoring, typically at the urban scale. 
We suppose to consider an urban area with some fixed reference 
stations (in general only a few) measuring some pollutants 
together with a network of low-cost sensors (in general numerous or at least more dense), cheaper but of lower quality. However, these sensors require calibration. Some of the reasons are that their performances can change in time or when they are moved (see \cite{BorregoAssessment2016,CastellLocalized2018}), or that they are sensitive to environmental factors (see \cite{LiangCalibrating2021}).
Moreover, apart from their lower quality, 
most of the low-cost sensors do not provide concentration measurements but rather intensity or tension measurements (\cite{WangMetal2010}). 
Therefore, calibration models must be constructed for the sensors in order to interpret their measurements.

Measurements from a network of low-cost sensors can be corrected using pointwise methods, 
multiple linear regression models or more complicated ones, treating each sensor independently of the others, or using global methods, taking into account several sensors at once. 

\noindent 
A common pointwise method involves fitting, what we call, an \emph{imported} model. More precisely, first, a low-cost sensor is placed near a fixed reference station to fit a correction model. Then, the sensor is moved and the previous model is used at the new measurement site, i.e. the model is imported at the new location. If the sensor is not moved and the model is used at the reference station site, the model is referred to as a \emph{collocated} model. The literature on the statistical models used in this process is extensive, covering methods such as linear regression (\cite{AhumadaCalibration2022, WinterScalable2025, spinelle2015,HongLongTerm2021,DongCalibration2025}, generalized additive models (\cite{vanZoestCalibration2019}), and neural network approaches (\cite{AhumadaCalibration2022, OkaforImproving2020,KozielField2024,ElbestarAir2025}).
Finally, another pointwise model studied previously (see \cite{WinterScalable2025, BobbiaStatistical2025}) is the \emph{non-collocated} model, obtained by fitting the measurements of a suitably chosen reference station with the measures coming from a low-cost sensor located in a similar site.  
This has the advantage of not requiring the sensor to be collocated with a reference station, 
and has proved generally better for calibration than the imported model.

\noindent 
On the other hand, global methods lead to network calibration strategies. For example, the  \emph{multihop} procedure (\cite{MaagSurvey2018}), which consists in calibrating a first sensor, then using this calibrated sensor as a reference to calibrate the next, and so on. This is therefore a case of serial calibration, which can be carried out using mobile sensors, for example. 

\noindent The \emph{iterative correction} proposed by \cite{corr-ite2022} is also a global method. 
Its principle is to divide measurement sites into two networks, N$_1$ for reference stations and N$_2$ for low-cost sensors. 
At a given point in time, the data from the N$_2$ network are interpolated to the sites in the N$_1$ network, using ordinary kriging (\cite{MatheronTraite62, cressie1993}).  Since we have reliable measurements on the N$_1$ network. 
We can then calculate the residuals between measurements and interpolation. The residuals are then interpolated on the N$_2$ network by kriging. This produces corrected measurements for each low-cost sensor.  
This procedure may or not may be iterated, treating the corrected measurements as data on the N$_2$ network. 

In this article, we use geographically weighted regression (GWR) 
as an alternative method for calibrating or correcting low-cost sensors. 
GWR is a local spatial statistical technique that can be used to model real phenomena 
by incorporating spatial nonlinearities.

\noindent 
This method was introduced by \citet*{gwr:brunsdon1996} 
and has been used in many fields. 
In air pollution, \citet{gwr:pm-china} combined GWR and PCA to develop a model for estimating the annual concentration of PM$_{2.5}$ in a Chinese province.
For soil pollution, \citet{gwr:heavy-metals} used GWR to assess the relationship between the presence of heavy metals and human activity (land-use).  
\citet{gwr:water-pakistan} modeled the relationship between the price of drinking water and place of residence in Pakistan.

For each $s\in D$, a domain of space, GWR explains $Y(s)$ by a linear model of the form :
\begin{equation*}	
Y(s) = \beta_0(s) + \beta_1(s) X_1 (s) + \ldots \beta_p(s) X_p(s) + \epsilon(s)
\end{equation*}

\noindent 
In other words, it is a simple multiple linear model at each given point of the space, 
but as the explanatory variables and the parameters of this linear regression model depend on space, 
this formalism allows us to take into account any spatial nonlinearities of the phenomenon under study (see, for example, \cite{gwr:spatial-nonstat}). 
It should be noted that this model is more general than land-use regression models (see \cite{hoek2008review} for a review) allowing to assess spatial variation of a phenomena through the use of explanatory variables which are maps.

\noindent 
The additional flexibility provided by GWR 
allowing spatially varying coefficients, makes it possible to model the fact that by moving a low-cost sensor, the coefficients of its calibration model change according to position. 
In other words, low-cost sensor measurements can be impacted differently from one environment to another. 
This was highlighted for example in the previous study carried out by the authors in Rouen (see \cite{BobbiaStatistical2025}). 

In this article, we illustrate the application of  calibration methods based on the GWR paradigm 
to the so-called Antwerp dataset (\cite{Dataset_Anvers, Data_Anvers}) 
proposed by the Joint Research Center to study AirSensEUR low-cost sensors
(\cite{ASE_TechA, ASE_TechB}). 
Antwerp exhibits a very dense situation in terms of reference and low-cost sensors networks with 9 reference stations, 12 collocated sensors and 22 non-collocated low-cost sensors.

The paper is organized as follows. Section 2 provides materials and methods. We first present the study area, the low-cost sensors and reference analyzers of the SensEURCity dataset in Antwerp, the data preprocessing.
Then, the structure of the Geographically Weighted Regression (GWR) model is presented together with model estimation and hyper-parameters selection. Section 3 is dedicated to the low-cost sensors calibration proposal together with the validation scheme. 
Section 4 addresses calibration results for NO$_2$ on the Antwerp dataset, together with some remarks about the estimated GWR model, the spatial character of estimated coefficients, opening the the possibility to calibrate non-collocated low-cost sensors.
Section 5 contains a short conclusion and discussion.

\section{MATERIALS AND METHODS}

\subsection{Data}
\label{ssec:data}

This section presents the dataset of measurements used in this study. 
This dataset has been made available by the Joint Research Center to study AirSensEUR sensors in three European cities (Antwerp, Zagreb and Oslo), see \cite{Dataset_Anvers, Data_Anvers}. 
In each city, AirSensEUR sensors have been deployed and measured  NO$_2$, NO, CO, PM$_{10}$, PM$_{2.5}$ and PM$_1$. 
Measurements of gaseous pollutants are expressed in nA, 
while particulate matter measurements are concentrations in \mgm3. 
Sensors also measure temperature in \celsius C, relative humidity in \% and atmospheric pressure in mbar. 
More specifically, the study focused on data from Antwerp.

\subsubsection{SensEURCity dataset in Antwerp}

\begin{figure}[H]
	\centering
	\begin{tabular}{cc}
			\includegraphics[height = 6cm]{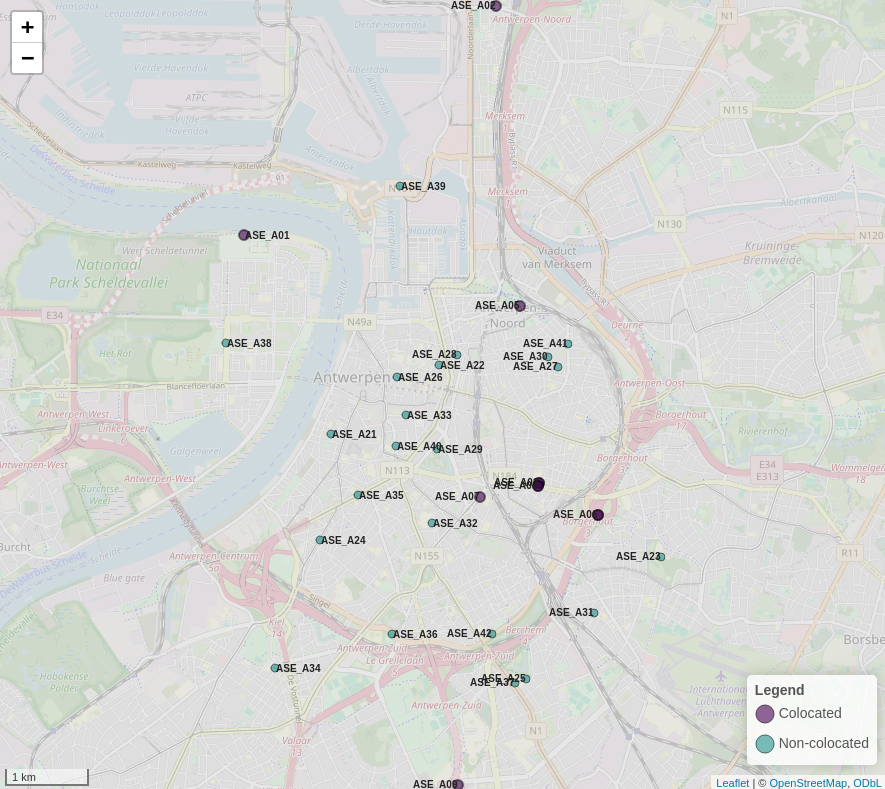} &
			\includegraphics[height = 6cm]{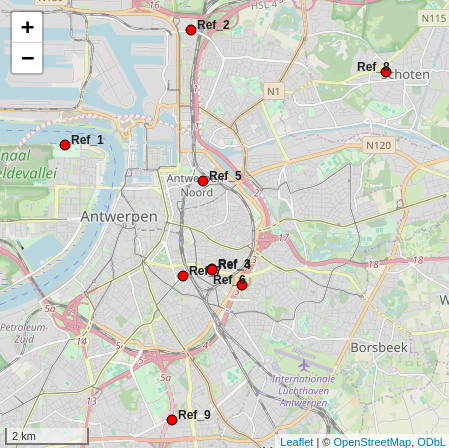}\\
	\end{tabular}
	\caption{Map of Antwerp : position  sensor and station locations during the deployment phase.}
	\label{fig:map_sensors}
\end{figure}

The Antwerp dataset comprises 9 reference stations and 34 low-cost sensors, all of which were collocated and deployed within the city. 
To increase the readability, we have renamed the sites of the reference stations (going from Ref\_1 to Ref\_9) and the low-cost sensors. The correspondence with the original names of the devices defined in \cite{Data_Anvers} is given in the \hyperref[app:nom-sensors]{Appendix}.
It consists of measurements carried out in three phases.
During Phase 1 (P1), the 34 sensors were located at a single station, Ref\_3. The site of the reference station Ref\_3 corresponds, on the map in Figure \ref{fig:map_sensors}, to that of sensors ASE\_A03 and ASE\_A13. 
The 34 sensors were installed at station Ref\_3 between 03/26/2020 and 04/03/2020, 
then removed between 06/15/2020 and 06/18/2020. 
The sensors were then moved to their deployment sites, corresponding to phase 2 (P2).
Twelve sensors were positioned in the vicinity of 9 reference stations, 
while the remaining 22 were deployed throughout the city of Antwerp. 
This phase lasted around 8 months. 
The sensors were deployed between 06/15/2020 and 06/29/2020. 
They were removed between 02/17/2021 and 02/26/2021. 
Finally, all 34 low-cost sensors were again collocated at the Ref\_3 station during a third phase (P3) lasting one and a half months.  This period begins at the end of the previous one and ends between April 13 and 15, 2021.

\noindent 
Note that few sensors are located in areas close to green spaces. 
The few sensors for which this is not the case are ASE\_A05, ASE\_A23, ASE\_A25, ASE\_A29 (background sites), ASE\_A38 and ASE\_A42 (traffic sites).
Similarly, few sensors are located directly on major roads. 
These are ASE\_A06, ASE\_A09, ASE\_A16, ASE\_A36, ASE\_A37, ASE\_A39 and ASE\_A42, which are traffic sensors. ASE\_09 is an exception, as the sensor is located on a road but away from the city. 
Table \ref{tab:type_loc} contains the characterization of sensors' locations (urban or suburban versus traffic, industry or background) and in bold those of low-cost sensors collocated, 2 of each type.

\begin{table}[H]
    \centering
    \begin{tabular}{|c|>{\centering\arraybackslash}m{40mm}|>{\centering\arraybackslash}m{20mm}|>{\centering\arraybackslash}m{40mm}|}
    \hline
         & Traffic & Industry & Background \\
    \hline
        Urban & \textbf{A04, A14, A06, A16}, A24, A26, A35, A36, A37, A38, A40, A41, A42 & \textbf{A01, A02} & \textbf{A03, A13, A05}, A21, A22, A27, A28, A29, A30, A31, A32, A33 \\
    \hline
        Suburban & \textbf{A07}, A39 & None. & \textbf{A08, A09}, A23, A25, A34\\
    \hline
    \end{tabular}
    \caption{Characterization of sensors' locations. In bold, the collocated low-cost sensors}
    \label{tab:type_loc}
\end{table}

\subsubsection{Data preprocessing}

Raw measurements of sensors are sampled every minute. 
However, we decided to aggregate the measurements to obtain an hourly average, 
as this is the most common method for examining reference measurements. 
Also, most of the exceedance thresholds defined by European regulation are expressed on an hourly basis 
(see for example \cite*{eea-no2}) and hourly measurement is commonly studied in the literature (e.g. \cite*{WinterScalable2025}).

\noindent 
To transform the raw data (available every minute) into hourly averages, 
we applied the following procedure, the same one as for the AirSensEUR study in Rouen (see \cite{BobbiaStatistical2025}) to each sensor:

\begin{enumerate} 
\item Remove outliers from the dataset.
This was done using the flags added by the Joint Research Center with the data.
\item Aggregate minute data into quarter-hour data. If at least 75\% of the period was available, the quarter-hour was retained. If not, it was considered unavailable, as the data was not sufficiently representative of the period. This corresponds to a minimum of 12 minutes in the quarter-hour.
\item Aggregate quarter-hour data into hourly data. An hour is considered sufficiently representative if 3 or 4 quarter-hours are available. Otherwise, the hour is deleted from the dataset.
\end{enumerate}

Our study focuses on nitrogen dioxide, 
and we have included, on one hand, NO and CO for cross-sensitivity
and, on the other hand, meteorological parameters (internal temperature, relative humidity).
For convenience, the NO$_2$ reference data initially provided in ppb 
have been transformed into \mgm3 using the formula taken from \cite{seinfeld2016atmospheric} 
and derived from the law of perfect gases\footnote{$[\text{\mgm3}] = \frac{[ppb]\times P \times M}{1000\times R\times T}$ where $P$ represents atmospheric pressure in $Pa$, $T$ is the thermodynamic temperature in kelvins, $M$ the molar mass in g/mol and $R$ is the universal constant of perfect gases.}.

\subsection{Geographically Weighted Regression (GWR)}

Geographically weighted regression (GWR), as described in \citet{gwr:brunsdon1996}, enables real phenomena to be modeled using a linear model whose coefficients and covariates depend on spatial location. 
It can therefore incorporate spatial nonlinearities (see for example, \cite{gwr:spatial-nonstat}). 

\subsubsection{Model structure}

In this section, we briefly summarize the principle of the method. 
Let $D$ be a compact subspace of $\R^2$,  
$Y(s)$ be the real random variable of interest associated with the position $s \in D$, 
and let $X_1(s), X_2(s), \ldots, X_p(s)$ be the
$p$ real random variables (the covariates). 
For each $s \in D$,   $Y(s)$ is modeled by a
linear model of the form:
\begin{equation}
	Y(s) = \beta_0(s) + \beta_1(s) X_1 (s) + \ldots \beta_p(s) X_p(s) + \epsilon(s)
	\label{eq:gwr}
\end{equation}
where $\beta_0(s)$ is the value of the intercept at position $s$ and
$\beta_j(s)$ ($j = 1, \ldots, p$) are the coefficients of the regressors, functions of position $s$.
Here $\epsilon(s)$ is the error term, with zero mean and variance $\sigma^2$ assumed to be constant over $D$. 
Moreover, $\epsilon(s)$ is assumed to be independent of
$\epsilon(s^\prime)$ for all $s\neq s^\prime$.

\subsubsection{Model estimation and hyper-parameters}

To estimate the parameter vector
$\beta(s) = (\beta_0(s), \beta_1(s), \ldots, \beta_p(s))^T \in \bkR^{p+1}$ 
in each position $s\in D$, 
we have the measurements
of the processes $X_1, \ldots, X_p$ and $Y$ in only $q$ distinct points of $D$, denoted by $s_1, s_2, \ldots s_q$.
The GWR method consists in estimating $\beta(s)$ by minimizing the following weighted least squares criterion:
\begin{equation}
G(\beta(s)) = \sum_{k = 1}^q w(s-s_k) 
\pa{Y(s_k) - \beta_0(s) - \beta_1(s) X_1(s_k) - \ldots - \beta_p(s) X_p(s_k)}^2 
\end{equation}
where $w : \bkR^2 \rightarrow \bkR^+$ 
is a weight function such that there exist a unique solution to the minimization problem at any point $s\in D$.
The function $w$ reaches its maximum in 0, which means that for a given $k$, $w(s-s_k)$ is maximal if $s=s_k$.
Obviously, the choice of $w$ is important and, for example, \citet{gwr:brunsdon1996} suggest taking:
\begin{equation}
	w(s-s_k) = exp\pa{-\lambda\,\norm{s-s_k}^2}
\end{equation}
where $\lambda > 0$ is a constant specified by the user according to the problem.

\noindent 
Noting $\Ybf = (Y(s_1), Y(s_2), \ldots, Y(s_q))^T$, $\Xbf = (\mathds{1}_q,
\Xbf_1, \Xbf_2, \ldots, \Xbf_p)$ with for all $j=1, 2, \ldots, p$, $\Xbf_j =
(X_j(s_1), X_j(s_2), \ldots, X_j(s_q))^T$ and $W(s) = diag(w(s-s_1), w(s-s_2),
\ldots, w(s-s_q))$.
The matrix $W(s)$ is diagonal of size $q \times
q$ and we can rewrite the function $G(\beta(s))$ as:
\begin{equation}
	G(\beta(s)) = (\Ybf - \Xbf \beta(s))^T W(s) (\Ybf - \Xbf \beta(s)).
\end{equation}
The solution to the minimization problem is, for $s\in D$ such that the matrix $\Xbf^T W(s) \Xbf$ is positive definite, given by:
\begin{equation}
\wbeta(s) = \pa{\Xbf^T W(s) \Xbf}^{-1} \Xbf^T W(s)  \Ybf.
\end{equation}

\noindent 
In particular, this expression shows the importance of the choice of $w$, since depending on $w$ and $s$, the matrix $\Xbf^T W(s) \Xbf$ may not be invertible. 

\subsubsection{GWR with repeated measurements}

The following remark addresses the use of the GWR method when
repeated measurements are available.

If for all $k=1, \ldots, q$, we have
of $n_k \geq 1$ observations of the vector 
$$(X_1(s_k), X_2(s_k), \ldots,
X_p(s_k), Y(s_k)),$$ denoted by $$(X_{1,i}(s_k), X_{2,i}(s_k), \ldots,
X_{p,i}(s_k), Y_i(s_k))_{1 \leq i \leq n_k},$$ 
then the criterion to minimize becomes:
$$G(\beta(s)) = \sum_{k=1}^q \sum_{i=1}^{n_k} w(s-s_k)
\pa{Y_i(s_k) - \beta_0(s_k) - \beta_1(s_k) X_{1,i}(s_k) - \ldots - \beta_p(s_k) X_{p,i}(s_k)}^2$$
Noting,
\begin{align*}
	\Ybftild &= (Y_1(s_1), \ldots, Y_{n_1}(s_1), Y_1(s_2), \ldots, Y_{n_2}(s_2),
	\ldots, Y_1(s_q), \ldots, Y_{n_q}(s_q))^T \\
\Xbftild^T &= \begin{pmatrix}
1 & \ldots & 1 & \ldots & 1 & \ldots & 1 \\
X_{1,1}(s_1) & \ldots & X_{1,n_1}(s_1) & \ldots & X_{1,1}(s_q) & \ldots & X_{1,n_q}(s_q) \\ 
\vdots & & \vdots & & \vdots & & \vdots \\
X_{p,1}(s_1) & \ldots & X_{p,n_1}(s_1) & \ldots & X_{p,1}(s_q) & \ldots & X_{p,n_q}(s_q) \\ 
\end{pmatrix} \\
	\Wtild(s) &= \diag(\underbrace{w(s-s_1), \ldots, w(s-s_1)}_{n_1}, \ldots,
	\underbrace{w(s-s_q), \ldots, w(s-s_q)}_{n_q})
\end{align*}
we can rewrite the previous weighted least squares criterion as:
$$G(\beta(s)) = (\Ybftild - \Xbftild \beta(s))^T \Wtild(s) (\Ybftild - \Xbftild \beta(s))$$
leading to the solution : $\wbeta(s) = \pa{\Xbftild^T
\Wtild(s) \Xbftild}^{-1} \Xbftild^T \Wtild(s) \Ybftild$. 
For sure, this solution exist if and only if the matrix $\Xbftild^T
\Wtild(s) \Xbftild$ is invertible.

\noindent It is then possible to obtain a solution, even if the  number of measurements at each point $s_k$ is not the same, which may be the case in our problem (in the event of sensor failure, for example, or activation at different dates).
In addition, it should be noted that a drawback of this method is that it requires observations of the dependent variable and its covariates to be located at the same $s_k$ points.

The GWR method and all subsequent numerical results were obtained using a direct implementation in \texttt{R} (\cite{R}).
Multiple packages are available in \texttt{R} to implement GWR: see for example \texttt{spgwr} \citep{spgwr}, \texttt{gwrr} \citep{gwrr} or \texttt{GWmodel} \citep{GWmodel}.

\section{LOW-COST SENSOR CALIBRATION PROPOSAL}

\subsection{GWR for network calibration}

We examine in this section, how to use GWR for calibration purposes. 
Several ways are available to fit the model requirements depending on the actual constraints.

Consider a monitoring network made up of low-cost sensors and reference stations 
and assume that there are $q$ reference stations to which $q$ low-cost sensors are collocated. 
Let's denote $s_1, s_2, \ldots, s_q$ the positions of these stations, 
and $\mathcal{S} = \{s_1, s_2, \ldots, s_q\}$. 
In addition, we have $K$ low-cost sensors, not collocated with a reference station 
and whose positions are noted $z_1, z_2, \ldots, z_K$ 
and $\mathcal{Z} = \{z_1, z_2, \ldots, z_K\}$.

\noindent 
Using a GWR model,
we want  to correct the measurements of these low-cost sensors 
located at positions $z_1, \ldots, z_K$. 
Ideally, the variable to be explained in the GWR model would be the actual concentration, 
but this is unknown. 
However, since reference stations are precise instruments, providing reliable, high-quality concentration measurements, they can be used instead of actual concentrations. 
Let us denote $Y(s_j)$ the measurement made by the reference station at point $s_j$, for $j\in \{1\ldots q \}$. 
In addition, let us denote $P$ the measurement of the pollutant of interest provided by the sensor 
and $X_1, \ldots, X_p$ the available covariates,
 either provided by the sensors or by another source. 
 It is assumed that measurements of variables $P, X_1, \ldots, X_p$
are available at all points $s \in \mathcal{S} \bigcup \mathcal{Z}$.

\noindent 
This framework enables us to calibrate and/or correct the low-cost sensors located at the $\mathcal{Z}$ points 
using GWR model 
and the measurements of variables  $P, X_1, \ldots, X_p, Y$ at the $\mathcal{S}$ points.

The coefficients of the GWR model can be estimated by using only the measurements observed at the time in question, or by using a history of measurements. 
The observation of $P(s)$ at time $t$ will hereafter be referred to as $P^t(s)$. 
This notation will be extended of all the variables.

For each low-cost sensor located at a point $z_k$ ($k=1, \ldots, K$),  
we propose to correct the measurement at time $t$, $P^t(z_k)$, 
by the value
$P^t_{Cor}(z_k)$ defined by :
$$P^t_{Cor}(z_k)= \wbeta_0(z_k) + \wbeta_1(z_k) X^t_1(z_k) + \ldots +
\wbeta_p(z_k) X^t_p(z_k) + \wbeta_{MS}(z_k) P^t(z_k)$$
where the vector 
$\wbeta(z_k) = (\wbeta_0(z_k), \ldots, \wbeta_p(z_k), \wbeta_{MS}(z_k))^T$ 
is the solution of the following weighted least squares problem:
\begin{equation*}
\wbeta(z_k) = \underset{\beta\in\R^{p+2}}{\arg\min}\; G(\beta,z_k)
\end{equation*}
with
\begin{equation}
G(\beta,z_k) = \sum_{t = 1}^n \sum_{j = 1}^q w(z_k-s_j) \pa{Y^t(s_j) - \beta_0 -
\beta_1 X^t_1(s_j) - \ldots - \beta_p X^t_p(s_j) - \beta_{MS} P^t(s_j)}^2.
\end{equation}

\subsubsection{SGWR: GWR on standardized data}

As mentioned in Section \ref{ssec:data}, low-cost sensors' measurements of gaseous pollutants are expressed in nA instead of usual concentration units (\mgm3 or ppb). 
It can be noticed from the data that the levels of amperage are different from one sensor to another, even when the environment and pollutants' concentrations are similar. 
To reduce the impact of the heterogeneity, we propose to introduce SGWR, a GWR model based on standardized data and defined by:

\begin{enumerate} 
\item 
Measurements made by each sensor are standardized. 
Thus, for each sensor and each variable, new data are of zero mean and unit variance. 
The standardization coefficients (mean $\mu_{i,j}$ and standard error $\sigma_{i,j}$ of each variable $i$ of each sensor $j$) are stored.
\item 
The GWR model is trained on standardized data, applying the classical  procedure detailed previously. 
\item 
Coefficients obtained at locations $s_j$ are then de-standardized, using the formula in Equation \eqref{eq:sgwr} 
where $\tilde{\beta}_i(s_j)$ denotes the $i$-th parameter estimated with standardized data:

\begin{equation}
    \begin{cases}
        \beta_i(s_j) = \tilde{\beta}_i(s_j) / \sigma_{i,j} & \text{ if } i\neq 0\\
        \beta_0(s_j) = \tilde{\beta}_0(s_j) - \sum_{i=1}^p \tilde{\beta}_i(s_j)\frac{\mu_{i,j}}{\sigma_{i,j}} & \\
    \end{cases}
    \label{eq:sgwr}
\end{equation}

Corrected measurements are then computed using the $\beta_i(s_j)$ and the original variables.
\end{enumerate} 

\subsubsection{Which variables to use?}

Explanatory variables of the calibration model must include 
the measurement of the pollutant of interest for the microsensor, 
since it is to be corrected. 
For $t \in \{1\ldots n\}$, $P^t(s)$ is the $t-$th measurement of the low-cost sensor placed at location $s$. 
Other measurements provided by the low-cost sensors: measurements of gaseous pollutants (NO, CO) and meteorological parameters (relative humidity and temperature) are also to be considered. 
We will designate $X^t_1(s)\ldots X^t_p(s)$ the observations at time $t$ of these $p = 4$ quantities measured by the microsensor at site $s$. 
We then propose a model of the form (using notations similar to those used in Equation \eqref{eq:gwr}):
\begin{equation}
	 P^t_{Cor}(s) = \beta_0(s) + \beta_1(s) X^t_1(s) +
	 \ldots + \beta_4(s) X^t_4(s) + \beta_{M}(s) P^t(s) 
	\label{eq:gwr10}
\end{equation}

\noindent 
The model defined by Equation \eqref{eq:gwr10} makes use of 5 variables (NO$_2$ measurement plus 4 other measurements) and will be called model GWR$_5$.

\subsubsection{Which models and competitors to consider?}

We will then compare four models which include measurements of gaseous pollutants (NO, NO$_2$, CO), humidity and temperature as covariates. 
They consist in a GWR model, a SGWR model, a linear collocated model which 
is \textit{a priori} the most efficient model and a non-collocated linear model. 
These two last models are defined by Equation \eqref{eq:gwr10} 
removing the dependence on $s$ and choosing accordingly the training observations. More precisely, the non-collocated model is constructed using $Y_{ref}$ observations from a reference station located elsewhere than the sensor location, and making sure to consider a station at a site with the same typology (see Table \ref{tab:type_loc} to identify the similar typologies).

As mentioned previously, other competitors could have been considered, for example the imported model. Based on conclusions from \cite{WinterScalable2025, BobbiaStatistical2025}, we have decided to compare the GWR and SGWR models to two models that can be estimated using the same time period in the training dataset, and compare methods that do not require collocation of all sensors.

\subsubsection{Which weight functions to use?}

The weight function is an important hyperparameter for fitting a GWR model. 

\begin{figure}[H]
	\centering
	\includegraphics{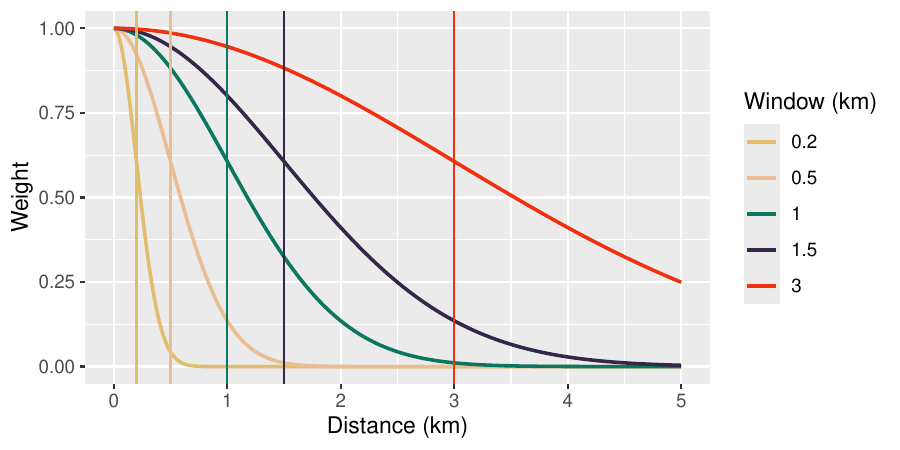}
	\caption{Weight functions for different values of window $B$}
	\label{fig:poids1}
\end{figure}

Noting $w_j(s)$ the weight applied to observations located at position $s_j$, when building the model at point $s$, we propose to apply the following Gaussian weight to build the GWR models:
\begin{equation}
	w_j(s) = \exp\left(-\frac{1}{2}\frac{\lVert s - s_j\rVert^2_2}{B^2}\right)
	\label{eq:poids1}
\end{equation}
\noindent 
where $B$ is a window that represents the distance for which $w(s) = \exp(-0.5) \approx 0.60$.
The smaller $B$ is, the faster the weights decrease, and the more emphasis is placed on observations at site $s_j$ that are spatially close to $s$. 
Figure \ref{fig:poids1} shows the weight functions for different values of $B$.

\subsection{Validation scheme}

This section details how to select learning and test sets in order to evaluate and validate the results of the procedure.

\subsubsection{Learning and test samples}

The deployment period (phase P2) is artificially divided into three parts: $S_0$, $S_1$ and $S_2$.
Each of them is built by selecting a given number of days and gathering all the measurements made by each sensor at every hour of those days.
More precisely, $S_1$ will be the sample used to build both collocated (C.) and non-collocated (NC.) models. 
The days selected for this sample correspond to 1 day out of 8, which allows each day of the week to be fairly considered when building models, and consists in 12.8\% of the data.
Then, $S_1$ is built taking about 2 days out of each set of 3 consecutive days, which consists in 61.1\% of the period. 
This sample is used to estimate the parameters of the GWR models.
Finally, the $S_2$ sample, consisting of days not already selected, is used to test and compare the performance of the models. It represents 26.1\% of the period.

\begin{figure}[H]
	\centering
	\includegraphics[width = 0.8\textwidth]{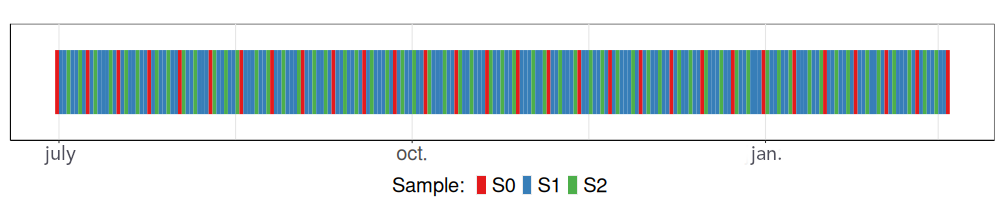}
	\caption{Timeline of the P2 period colored according to the distribution of days between the samples}
	\label{fig:samples_timeline}
\end{figure}

\noindent 
Figure \ref{fig:samples_timeline} presents the distribution of samples over the time period. 
The non-collocated models of each sensor are estimated using data from sample $S_1$.

Let us remark that in the sequel $S_0$ is in fact not used. Indeed, the so-called calibration model fulfills two functions for NO$_2$: converting measurements expressed in nA into concentrations in \mgm3, and performing the calibration itself. In our experiments, we considered various approaches to account for this double function. One approach was to fit  conversion models on $S_0$ and then to calibrate the converted measurements using GWR models fitted on $S_1$. However, this approach was no more efficient than SGWR and made evaluations of the GWR strategy more tricky due the cascade of models. Ultimately, we decided to omit  the results of this two-phase strategy from the paper to save space, leading to a set $S_0$ which is not used. Nevertheless, we left it in the experimental protocol in case someone might wish to reproduce this strategy.

Results are assessed on the test sample $S_2$. For each sensor at location $s$, we use the Root Mean Square Error (RMSE) and the percentage of explained variance (EV) defined respectively by:

\begin{equation}
\text{RMSE}(s) = \sqrt{\frac{1}{n}\,\sum_{t=1}^{n} (y_t(s) - \widehat{y}_t(s))^2}
\end{equation}
and  
\begin{equation}
\mbox{\rm EV}(s) = 1 - \frac{\sum_{t=1}^n (y_t(s) - \widehat y_t(s))^2}{\sum_{t=1}^n (y_t(s) - \overline{y}(s))^2}.
\end{equation}
where $y_t(s)$ is a reference measure, $\widehat y_t(s)$ is the predicted value and $n$ the sample size.

\subsubsection{Leave-one-out cross validation on reference stations}

To validate the GWR method for calibrating and/or correcting low-cost sensor measurements, 
the ideal would be to compare the corrections provided by the GWR model
at points $z_1, \ldots, z_K$ with the actual concentrations of the  pollutant which are not available. 
The only points where this is possible are the points $s_1, \ldots, s_q$
for which we have the measurements of the reference stations who are the closest to the real concentration. 
The natural strategy is therefore to implement a leave-one-out cross-validation on the reference stations. 
The principle is as follows. 
For each site $s$ of all sites $\mathcal{S} = \acc{s_1, s_2, \ldots, s_q}$ 
where a reference station and a sensor are collocated:

\begin{enumerate}
\item 
Only the observations of the training sample concerning the measurements of the sites $\mathcal{S}\backslash \acc{s}$ are considered, the measurements of the site $s$ are not taken into account in the construction of the model;
\item 
The parameters $\beta(s)$ are estimated using the GWR method;
\item Then, over the test period, the corrected values of the low-cost sensor located at the location $s$ are compared to the concentrations provided by the reference station of the same site.  
This will provide information on the expected performance for the low-cost sensors deployed at the sites $z_1, \ldots, z_K$. 
\end{enumerate}

\section{CALIBRATION RESULTS FOR NO$_2$}

This section presents the transformation/calibration models for low-cost sensors measurements obtained in Antwerp using GWR and SGWR models. 
We consider two different window choices. 
First, the window of 3000m, considered as large with respect to the city scale, 
and the adapted window of 1460m. 
This last window was obtained by calculating, in cross-validation and over the learning period, the RMSE as a function of the window, and taking the one leading to the minimum. 
Results are compared to those obtained with the collocated model (which is the ideal one but barely available in practice) and the non-collocated model (usually available in practice and easy to implement). 

Transformation models for NO$_2$ measurements in nA into concentrations in \mgm3 
are constructed using multiple covariates (gas measurements and meteorological parameters from sensors). 
This is necessary, as measures (intensity or tension) of the electrical current induced by NO$_2$ cannot fully explain NO$_2$ concentrations. 
This has been studied with collocated models (see \cite{BobbiaStatistical2025}) in a very similar context.

Let us mention that only 9 of the 12 collocated LCSs are considered, as 3 reference stations have 2 LCSs collocated at the same station. Only one of these are considered, to avoid over-representation of their location.

\subsection{Test performance of GWR}

We evaluate the performance of test set of the GWR type procedures using a first window related to the scale of the city, leading to a smooth version, and in a second section, an optimal window search by cross-validation on the learning set, more adapted. 

\subsubsection{Window 1: 3000m, a large window}
The percentage of explained variance EV of the different models over the S2 period,
are collected in  Table \ref{tab:ev-no2-all} and in Table \ref{tab:rmse-no2-all} 
the corresponding RMSE.

\begin{table}[H]
\centering
\begin{tabular}{l|c|c|c|c}
  \hline
 & C. & NC. & GWR$_5$ & SGWR$_5$ \\ 
  \hline
  ASE\_A01 & 66.00 & 43.30 & 25.70 & 34.20\\ 
  ASE\_A02 & 66.20 & 43.50 & 31.00 & 45.60\\ 
  ASE\_A03 & 71.10 & 66.70 & 50.50 & 42.30\\ 
  ASE\_A04 & 84.00 & 49.90 & 53.50 & 78.50\\ 
  ASE\_A05 & 50.80 & 48.50 & 35.40 & 40.10\\ 
  ASE\_A06 & 81.00 & 38.50 & 20.40 & 38.00\\ 
  ASE\_A07 & 71.10 & 67.00 & 49.40 & 65.90\\ 
  ASE\_A08 & 42.00 & 32.30 & 32.40 & 33.70\\ 
  ASE\_A09 & 74.40 & 62.10 & 47.10 & 57.80\\ 
   \hline
\end{tabular}
	\caption{Percentage of EV  for complete models over the test period (S2)}
	\label{tab:ev-no2-all}
\end{table}

\begin{table}[H]
\centering
\begin{tabular}{l|c|c|c|c}
  \hline
 & C. & NC. & GWR$_5$ & SGWR$_5$\\ 
  \hline
  ASE\_A01 & 8.00 & 10.40 & 11.90 & 11.20\\ 
  ASE\_A02 & 9.50 & 12.30 & 13.60 & 12.10\\ 
  ASE\_A03 & 9.10 & 9.80 & 11.90 & 12.90\\ 
  ASE\_A04 & 6.50 & 11.50 & 11.00 & 7.50\\ 
  ASE\_A05 & 10.60 & 10.80 & 12.10 & 11.70\\ 
  ASE\_A06 & 8.30 & 14.90 & 17.00 & 15.00\\ 
  ASE\_A07 & 8.20 & 8.70 & 10.80 & 8.90\\ 
  ASE\_A08 & 10.20 & 11.00 & 11.00 & 10.90\\ 
  ASE\_A09 & 6.80 & 8.20 & 9.70 & 8.70\\ 
   \hline
\end{tabular}
	\caption{RMSE for complete models over the test period (S2)}
	\label{tab:rmse-no2-all}
\end{table}

\noindent 
From the two tables, it appears that taking into account weather variables and measurements of other pollutants is sufficient to correctly transform NO$_2$ measurements. 
Indeed, the EVs of complete collocated models are between 42 and 84\%. 
It should be noted that they did not exceed 40\% with a simple model transforming linearly a microsensor measurement in nA into a corrected measurement in \mgm3 without any additional covariates 
(the detailed results are not reported here).

\noindent 
We observe moreover that there is a significant performance degradation when moving from the collocated model to the one on non-collocated data. 
The EV drops from 5 to 42 points in \%. 
The percentages of variance explained by the non-collocated model are between 32 and 67\% (RMSE between 8 and 15 \mgm3).

\noindent 
GWR$_5$ does not perform better than the non-collocated data model. 
Indeed, the EV (resp. RMSE) for each sensor are lower (resp. higher) for the model GWR$_5$. 
They are between 20\% and 53\% (respectively 9.7 and 17 \mgm3).

\noindent 
Finally, comparing GWR with the SGWR model leads to notice that 
SGWR$_5$ is more efficient than GWR$_5$ for 7 sensors out of 9. 
Their EV and RMSE are equal for the sensor ASE\_A08 and the only exception is ASE\_A03, with for SGWR$_5$ an EV of 42\% and for GWR$_5$ a 50\% EV. 
For all other sensors, we see an improvement in EV and RMSE by switching to a model on standardized data. 
In this case, EV increases from 1 to 26 points in \%. 
For five of the nine sensors, the improvement is at least 10 percentage points.

One of the conclusion of our previous work (see \cite{BobbiaStatistical2025}) 
shows that using several variables is a good approach to transform the measurements of low-cost sensors of NO$_2$ using multiple linear regression models. 
Moreover, it seems that among the GWR models, those on standardized data (SGWR) give the best results. 
These models are even competitive with the non-collocated model. All the results were presented with a given window defined according to geographical considerations, but whose value must be tuned by cross-validation.

\subsubsection{Window 2: 1460m, an adapted window}

An optimal window search by cross-validation on the learning set, leads to consider the value $B = 1460$m. 
We indicate in  Table \ref{tab:ev-no2-all2} the EV percentages of the different models over the period S2, and in Table \ref{tab:rmse-no2-all2} the associated RMSE. Since the collocated and non-collocated models are not influenced by the choice of window, the associated scores are the same as previously.

\begin{table}[H]
\centering
\begin{tabular}{l|c|c|c|c}
  \hline
    & C. & NC. & GWR$_5$ & SGWR$_5$\\ 
  \hline
	  ASE\_A01 & 66.00 & 43.30 & 42.80 & 58.90\\ 
	  ASE\_A02 & 66.20 & 43.50 & 57.60 & 65.50\\ 
	  ASE\_A03 & 71.10 & 66.70 & 46.90 & 37.30\\ 
	  ASE\_A04 & 84.00 & 49.90 & 53.70 & 78.70\\ 
	  ASE\_A05 & 50.80 & 48.50 & 41.40 & 43.70\\ 
	  ASE\_A06 & 81.00 & 38.50 & 25.80 & 50.20\\ 
	  ASE\_A07 & 71.10 & 67.00 & 48.70 & 65.00\\ 
	  ASE\_A08 & 42.00 & 32.30 & 43.70 & 44.60\\ 
	  ASE\_A09 & 74.40 & 62.10 & 53.30 & 74.30\\
   \hline
\end{tabular}
	\caption{Percentage of EV for complete models over the test period (S2)}
	\label{tab:ev-no2-all2}
\end{table}

\begin{table}[H]
\centering
\begin{tabular}{l|c|c|c|c}
  \hline
	  & C. & NC. & GWR$_5$ & SGWR$_5$\\ 
  \hline
	  ASE\_A01 & 8.00 & 10.40 & 10.40 & 8.80\\ 
	  ASE\_A02 & 9.50 & 12.30 & 10.70 & 9.60\\ 
	  ASE\_A03 & 9.10 & 9.80 & 12.40 & 13.40\\ 
	  ASE\_A04 & 6.50 & 11.50 & 11.00 & 7.50\\ 
	  ASE\_A05 & 10.60 & 10.80 & 11.50 & 11.30\\ 
	  ASE\_A06 & 8.30 & 14.90 & 16.40 & 13.40\\ 
	  ASE\_A07 & 8.20 & 8.70 & 10.90 & 9.00\\ 
	  ASE\_A08 & 10.20 & 11.00 & 10.00 & 9.90\\ 
	  ASE\_A09 & 6.80 & 8.20 & 9.10 & 6.80\\
   \hline
\end{tabular}
	\caption{RMSE of complete models over the test period (S2)}
	\label{tab:rmse-no2-all2}
\end{table}

\noindent 
The performance for the two values of the window are very different: GWR$_5$(3000) has an average EV on the sensors of 38\% while GWR$_5$(1460) has one of 46\%. 
In addition, the GWR$_5$(1460) model gives better estimates for all sensors except ASE\_A03, for which the EV was decreased by 2 points. 
In terms of RMSE, the switch to the new window has reduced the error up to 3 \mgm3 (sensor ASE\_A02). 
Similarly, the average EV on the sensors of the SGWR$_5$(3000) model was 48\%, that of SGWR$_5$(1460) is 58\%. For the ASE\_A03 sensor, the decrease in the window caused the EV to drop by 5 points; it increased it for all other sensors. The average RMSE of the SGWR$_5$(3000) model was 11 \mgm3, it is 10 \mgm3 for the SGWR$_5$(1460) model.

The reduction of the window size allows to improve the transformation of low-cost sensor measurements, 
and it is all the more obvious for isolated sensors (ASE\_A01, ASE\_A02 and ASE\_A09, see Figure \ref{fig:map_sensors}), 
for which the EV is increased by 20\%. 
This is explained by the fact that tightening the window will more force the isolated low-cost sensors to correct themselves with their own information. 
However, it should be noted that the choice of window was made through cross-validation, which indicates that this window is the most appropriate to estimate a transformation model on an isolated sensor that would not be connected to any fixed station.

\begin{figure}[htb]
    \centering
    \includegraphics[width=0.7\linewidth]{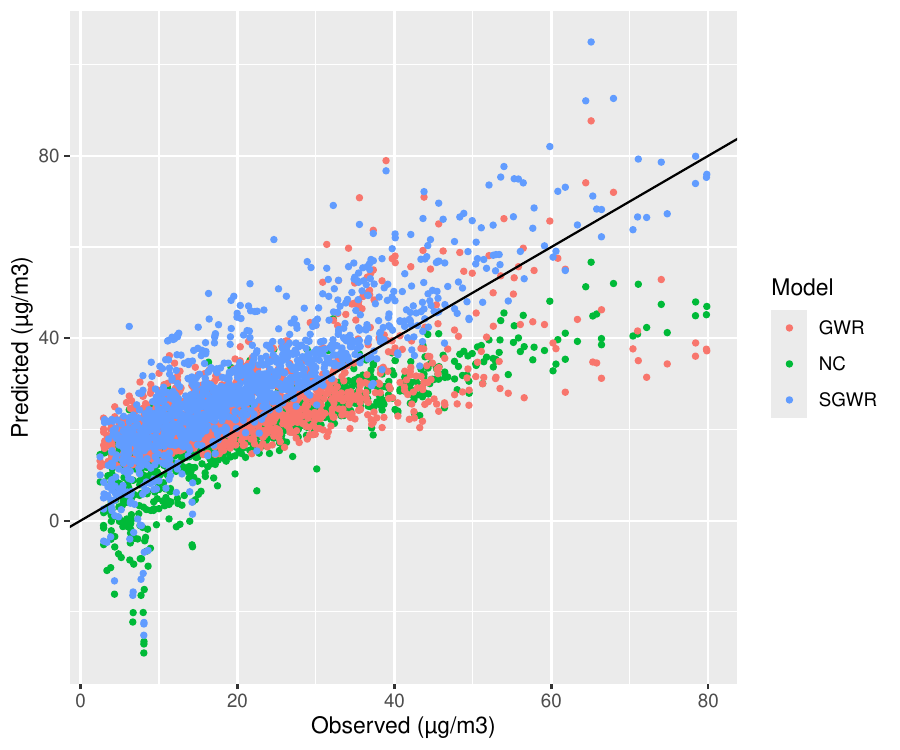}
    \caption{Scatterplot of measured NO$_2$ concentrations at Ref\_9 (x-axis) and corrected measurements from sensor ASE\_A09 (y-axis), depending on the method (NC, GWR or SGWR). The black line represents the curve $y=x$.}
    \label{fig:nuage_ASE9}
\end{figure}

Finally, Figure \ref{fig:nuage_ASE9} presents the scatterplot of the observed NO$_2$ concentrations at station Ref\_9 (x-axis), and the corrected measurements from sensor ASE\_A09 obtained using the three methods (y-axis). Overall, the three models tend to overestimate low NO$_2$ concentrations, with most points lying above the $y = x$ line. However, at higher concentrations (above 30 \mgm3), the non-collocated and GWR models tend to underestimate NO$_2$ levels. In contrast, the SGWR-corrected measurements are more consistent with the $y = x$ line and increase proportionally with the reference measurements. Overall, the SGWR model slightly overestimates NO$_2$ concentrations on the test set, indicating the presence of a small additive bias.

\subsection{Leave-one-out cross validation performance for GWR}

In order to evaluate the performance of the method at the points where there is no reference measurement, we propose to use a leave-one-out cross-validation. 
We will compare the proposed models considering only the window of 1460m. 
Table \ref{tab:cv-by-sensor}  presents the EV by sensor, on the left, and RMSE, on the right.

\begin{table}[H]
\centering
\footnotesize
\begin{tabular}{|l|c|c|c|}
  \hline
 Sensor & NC. & GWR$_5$ & SGWR$_5$ \\ 
  \hline
  ASE\_A01 & 43.30 & -6.70 & 29.20\\ 
  ASE\_A02 & 43.50 & -21.40 & 42.00\\ 
  ASE\_A03 & 66.70 & 20.70 & 30.90\\ 
  ASE\_A04 & 49.90 & 51.20 & 63.30\\ 
  ASE\_A05 & 48.50 & 22.70 & 21.50\\ 
  ASE\_A06 & 38.50 & 13.40 & 33.80\\ 
  ASE\_A07 & 67.00 & 46.30 & 62.50\\ 
  ASE\_A08 & 32.30 & 10.70 & 14.60\\ 
  ASE\_A09 & 62.10 & 38.80 & 28.30\\ 
   \hline
\end{tabular}\hfill
\begin{tabular}{|l|c|c|c|}
  \hline
  Sensor & NC. & GWR$_5$ & SGWR$_5$ \\ 
  \hline
  ASE\_A01 & 10.40 & 14.20 & 11.60\\ 
  ASE\_A02 & 12.30 & 18.10 & 12.50\\ 
  ASE\_A03 & 9.80 & 15.10 & 14.10\\ 
  ASE\_A04 & 11.50 & 11.30 & 9.80\\ 
  ASE\_A05 & 10.80 & 13.20 & 13.30\\ 
  ASE\_A06 & 14.90 & 17.70 & 15.50\\ 
  ASE\_A07 & 8.70 & 11.10 & 9.30\\ 
  ASE\_A08 & 11.00 & 12.60 & 12.30\\ 
  ASE\_A09 & 8.20 & 10.40 & 11.30\\ 
  \hline
  \bf Average & \bf 10.84 & \bf 13.74 & \bf 12.19\\
  \hline
\end{tabular}
\caption{Cross-validated EV (left) and RMSE (right) for complete models}
\label{tab:cv-by-sensor}
\end{table}

\noindent
On the right part of  Table \ref{tab:cv-by-sensor}, RMSE is better for SGWR 
and, more surprising, SGWR is often comparable to NC. This is remarkable. The cross-validated RMSE is defined by
\begin{equation}
    \text{CV RMSE} = \frac{1}{q}\sum_{j=1}^q \text{RMSE}_{\mathcal{S}\backslash\{s_j\}}
\end{equation}
where $\text{RMSE}_{\mathcal{S}\backslash\{s_j\}}$ represents the RMSE of the model constructed without making use of information from location $s_j$ and assessed for sensor at $s_j$. The CV RMSE is therefore estimated by 12.19 \mgm3, which is a good performance for a spatial CV RMSE.

\noindent 
The situation is less clear for EV (see the left part of Table \ref{tab:cv-by-sensor}). Let us examine in more detail the residuals (signed) for each low-cost sensor.

\begin{figure}[H]
    \centering
    \includegraphics[width=0.48\linewidth]{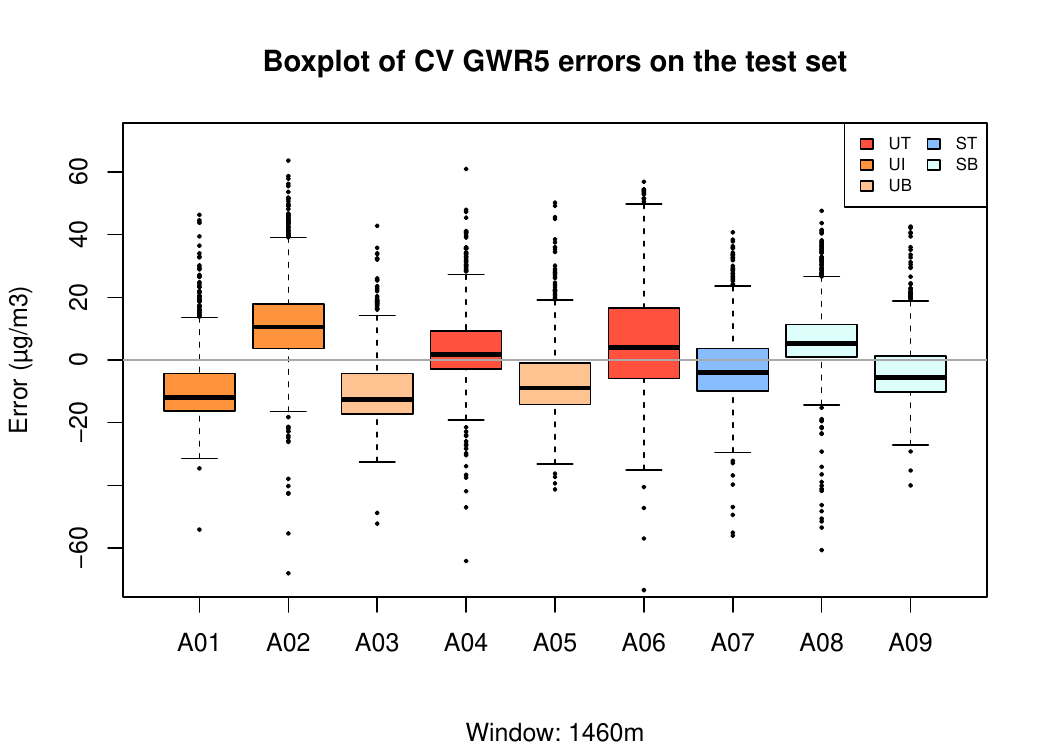}
    \includegraphics[width=0.48\linewidth]{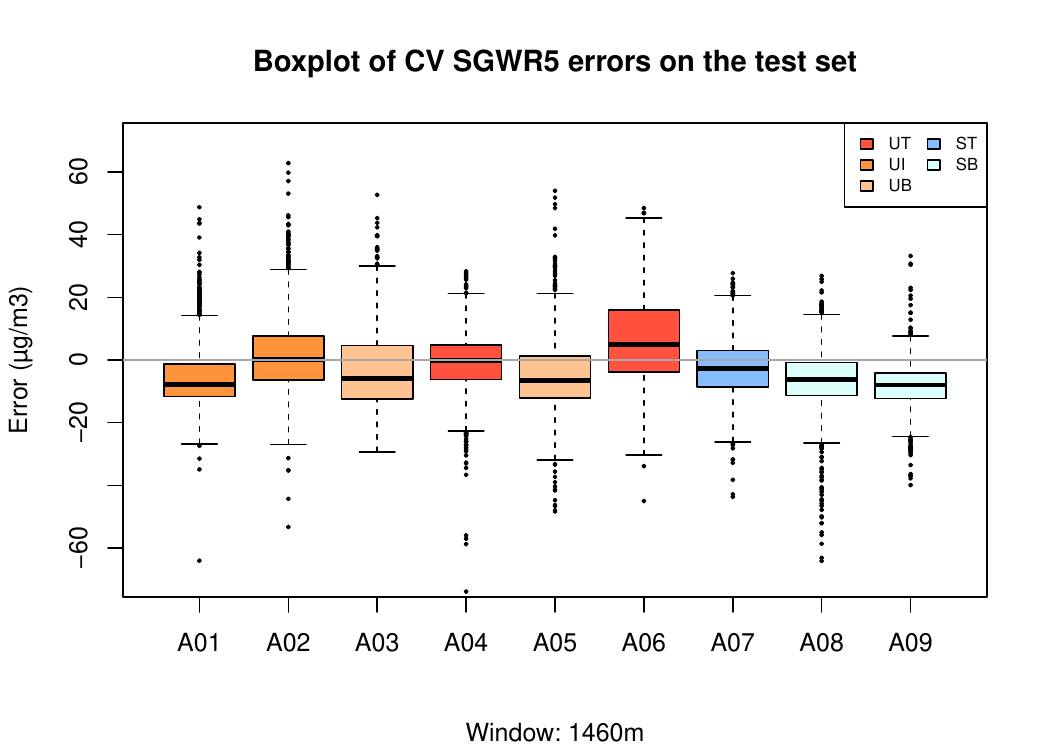}
    \caption{Boxplot of CV errors for GWR$_5$ and SGWR$_5$ models by sensor. Boxes are colored depending on the location type, as described in Table \ref{tab:type_loc}: red (Urban Traffic), orange (Urban Industrial), salmon (Urban Background), blue (Suburban Traffic), light blue (Suburban Background)} 
    \label{fig:boxplot_CVerrors}
\end{figure}

The boxplots of the cross-validated errors per station are displayed in Figure \ref{fig:boxplot_CVerrors}, for GWR on the left, and for SGWR on the right. The results are better for SGWR since all but a few boxes contains the value 0, while it is not the case for half of the stations for GWR. The boxplot for the traffic station ASE\_A06 is the largest one, because it is the one measuring the largest values with high variability.
Negative boxplots (measure greater than prediction) appear for ASE\_A08 and ASE\_A09, as expected since both are suburban background stations. The same phenomenon appears for ASE\_A01 which is an urban industrial station but not ASE\_A02. To end, for ASE\_A07, a suburban traffic station, leads to centered boxplot for GWR and SGWR.

To complement the analysis, it could be interesting to look at the influence of the hour of the day on the performance. For that, we define a metric that is the RMSE over space, at time $t$:
\begin{equation}
\text{RMSE}(t) = \sqrt{\frac{1}{q}\,\sum_{j=1}^{q} (y_t(s_j) - \widehat{y}_t(s_j))^2}
\end{equation}
where $y_t(s_j)$ is a reference measure, $\widehat y_t(s_j)$ is the predicted value and $q=9$ the number of collocated sensors.

\begin{figure}[H]
    \centering
    \includegraphics[width=\linewidth]{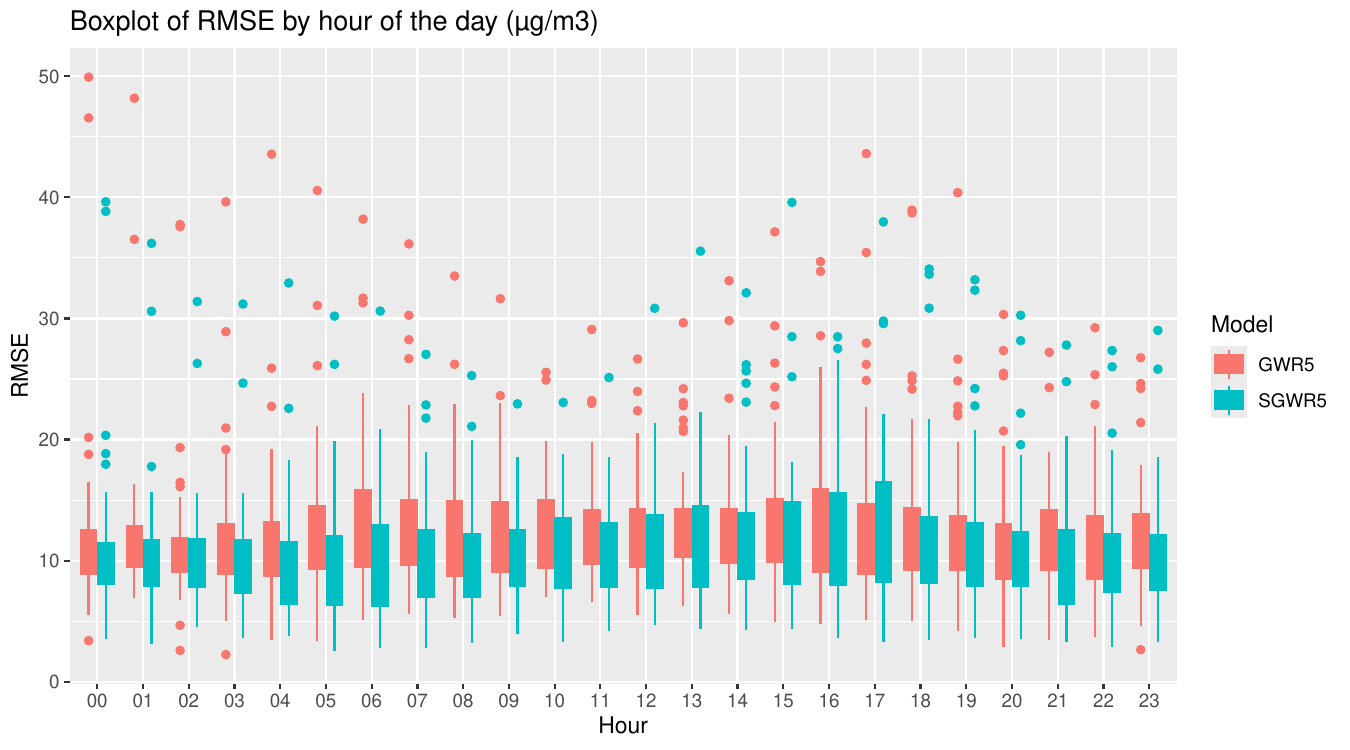}
    \caption{Boxplot of CV RMSE for GWR$_5$ and SGWR$_5$ models by hour of the day}
    \label{fig:boxplot_CV_RMSE_h}
\end{figure}

Figure \ref{fig:boxplot_CV_RMSE_h} shows the boxplots of CV RMSE for GWR$_5$ and SGWR$_5$ models by hour of the day. 
SGWR$_5$ outperforms GWR$_5$ and even if a small effect of the peak hours can be noticed, the distribution is very similar for all the hours, validating \textit{a posteriori} the choice of the learning set.

\subsection{Estimated spatial coefficients}

\begin{figure}[H]
    \centering
    \includegraphics[width=0.45\linewidth]{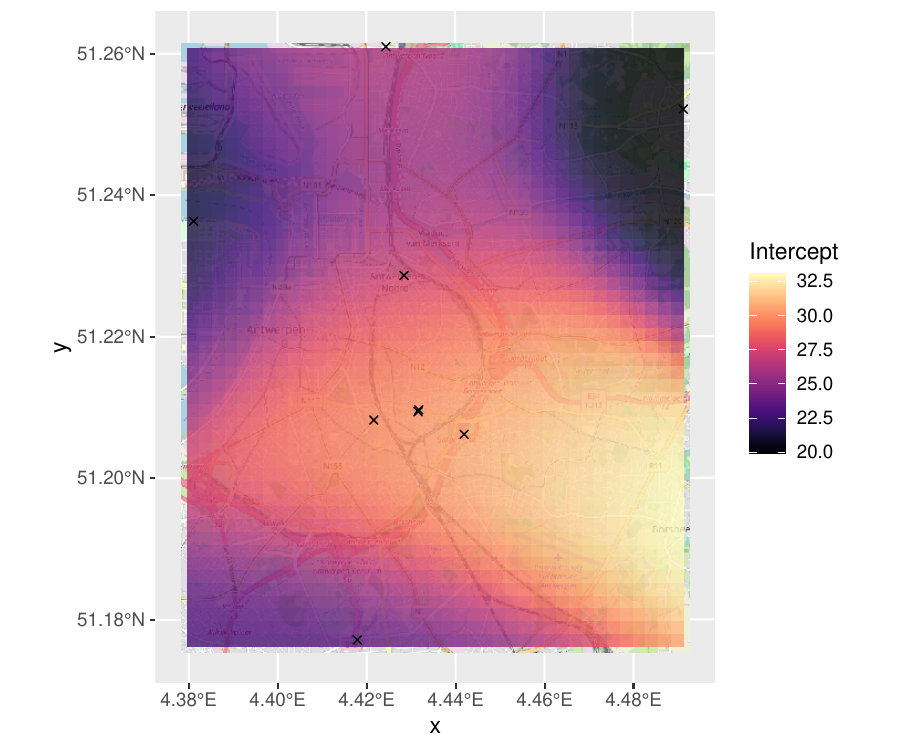}
    \includegraphics[width=0.45\linewidth]{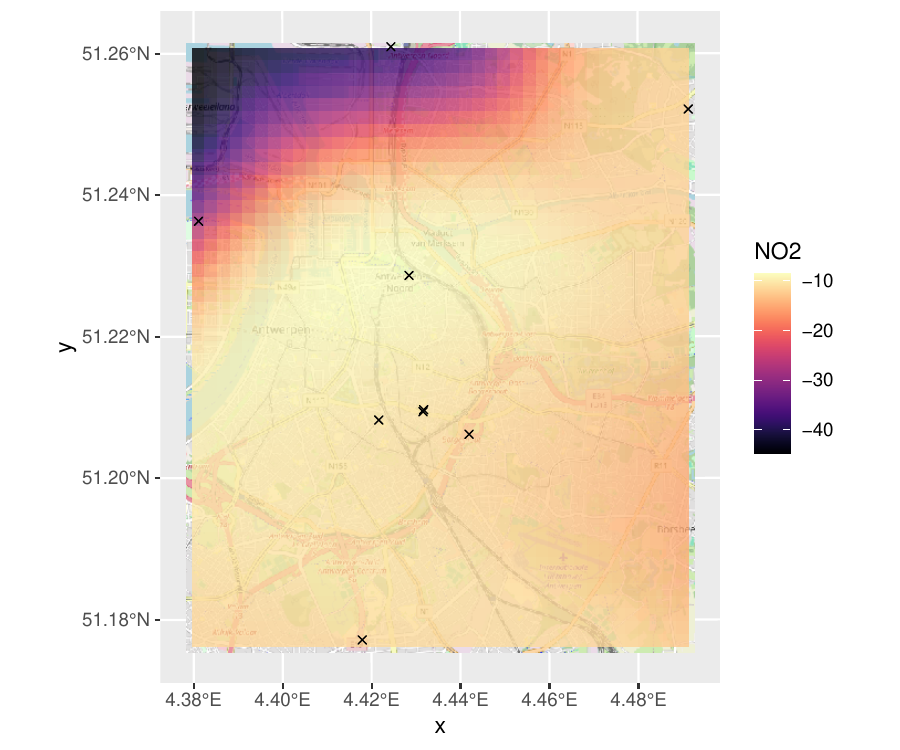}\\
    \includegraphics[width=0.45\linewidth]{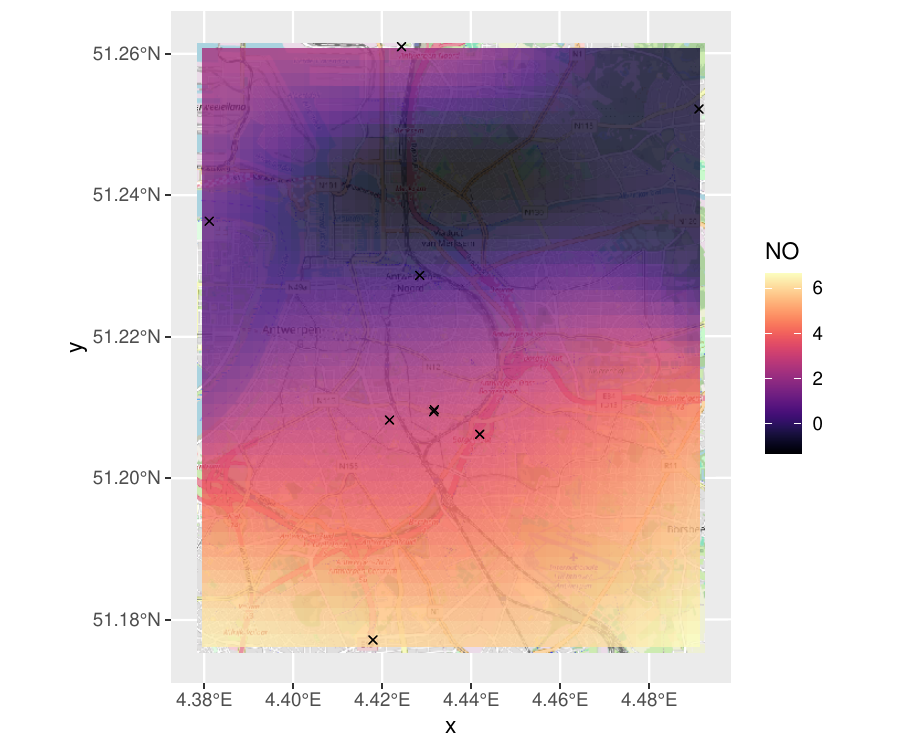}
    \includegraphics[width=0.45\linewidth]{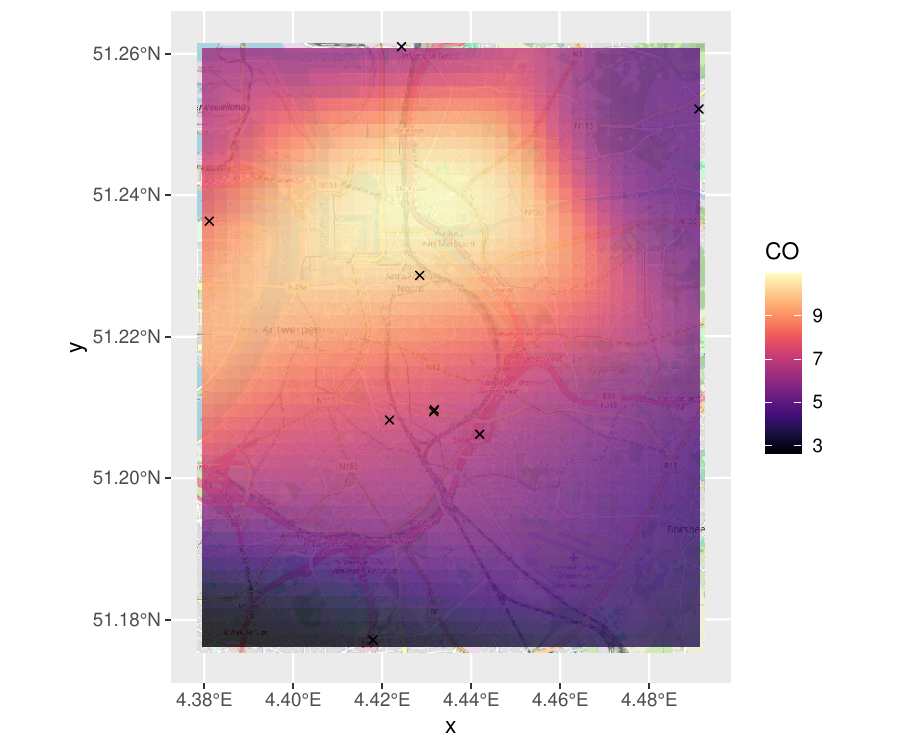}\\
    \includegraphics[width=0.45\linewidth]{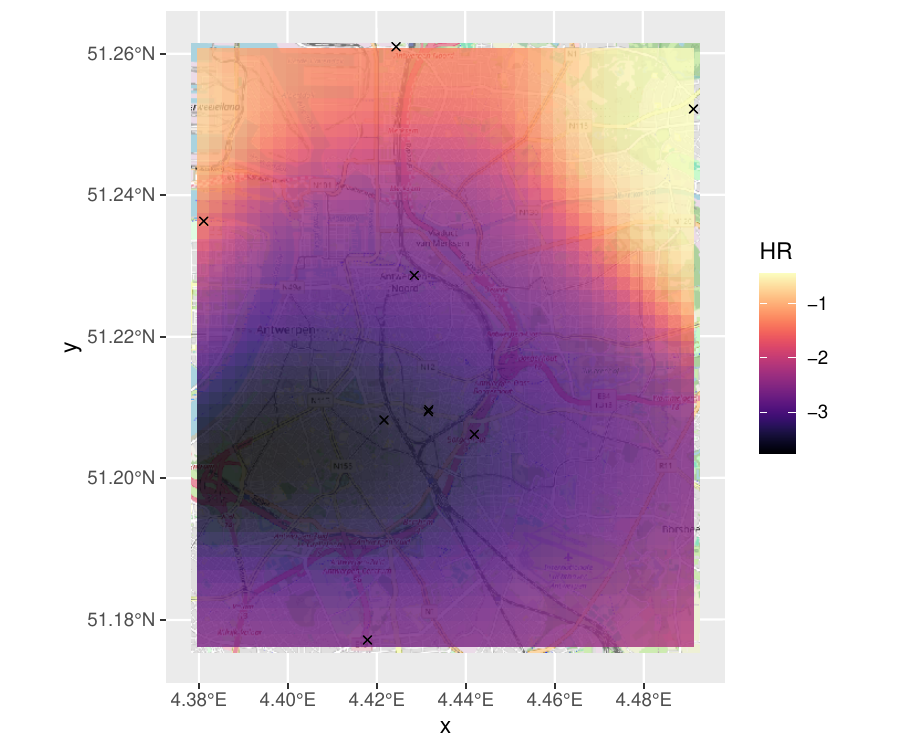}
    \includegraphics[width=0.45\linewidth]{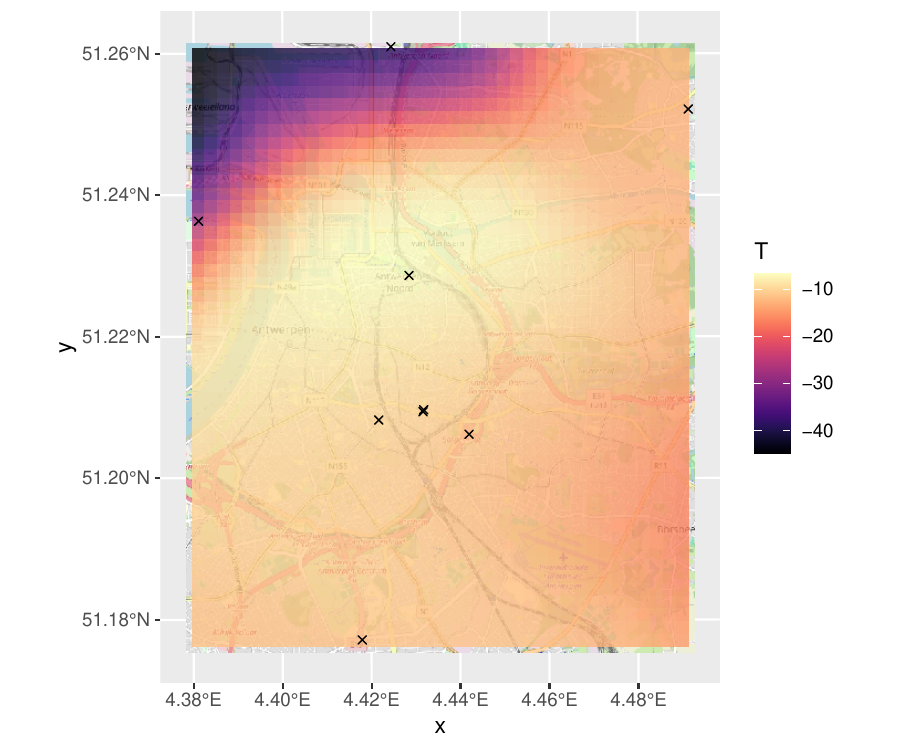}
    \caption{Map of coefficients $\beta(s)$ estimated by the SGWR model. Coefficients are expressed in \mgm3.}
    \label{fig:coeff-maps}
\end{figure}

The major contribution of the GWR approach lies in the spatial nature of the coefficients 
together with an unified view of several models. 
It means that there is a model for each point on the map and that we can visualize all of these models at a glance, just by looking at the maps.
The coefficient maps given by the Figure \ref{fig:coeff-maps} 
make it possible to use a sensor almost immediately after installation. 
The color of any point is the value of a coefficient.

It is possible to spatially interpret the coefficients by inspecting the maps with the one given by the map of the right part of Figure \ref{fig:map_sensors}. 
The points give the locations of the reference sites. 
For example, the map of the HR's coefficients (at the bottom left of the figure) is more negative inside the city while it is null or small at the edges of the map, as expected. 
Similarly, the intercept (map at the top left of the figure) is greater inside the city, which is supposed to be polluted, than at the corners of the map.

Even if it is possible to spatially interpret the coefficients, 
this assessment is limited. 
To explore if the GWR approach is reasonable for the calibration results on non-collocated sensors, 
we can check that the estimated coefficients of the collocated models (given by Table \ref{tab:coeff_collocated}) 
and the values of the GWR coefficient (given by Table \ref{tab:coeff_sgwr}), are compatible. As it can be seen, even if it is rather tedious, 
it is surprisingly (since we consider the collocated models) very often the case. For example the collocated models for ASE\_A01 and ASE\_A02 corresponding to the upper left corner of the map are quite different from the others and the signs vary accordingly with the GWR coefficients provided by the coefficient's maps and table. More precisely, for the sensor ASE\_02, the coefficients associated with NO$_2$ are estimated by -48.62 for the collocated model and by -38.08 for the SGWR model, while the estimated value of this coefficient is around -10 for sensors ASE\_03 to ASE\_A09 using both models. This leads to a promising prognosis for the calibration results on non-collocated sensors.

\begin{table}[H]
\centering
\begin{tabular}{lcccccc}
  \hline
 & Intercept & NO$_2$ & NO & CO & HR & T \\ 
  \hline
    ASE\_01 & 20.63 & -46.66 & 3.57 & 5.13 & 1.59 & -47.93 \\ 
  ASE\_02 & 26.00 & -48.62 & 2.86 & 2.37 & -0.58 & -51.15 \\ 
  ASE\_03 & 27.03 & -31.83 & 6.48 & 11.29 & -1.79 & -33.81 \\ 
  ASE\_04 & 29.11 & -14.13 & 1.90 & 2.91 & -4.03 & -11.84 \\ 
  ASE\_05 & 24.37 & 0.50 & -0.79 & 9.85 & -1.42 & 0.11 \\ 
  ASE\_06 & 33.58 & -23.55 & 7.72 & 2.04 & 0.03 & -27.61 \\ 
  ASE\_07 & 27.66 & -14.33 & 7.30 & 4.17 & -2.51 & -17.38 \\ 
  ASE\_08 & 19.07 & -10.69 & 0.08 & 3.93 & 1.91 & -11.88 \\ 
  ASE\_09 & 21.86 & -11.83 & 5.67 & 3.00 & -1.20 & -11.37 \\ 
   \hline
\end{tabular}
\caption{Coefficients estimated for collocated models on standardized values}
\label{tab:coeff_collocated}
\end{table}

\begin{table}[H]
    \centering
    \begin{tabular}{lcccccc}
    \hline
      & Intercept & NO$_2$ & NO & CO & HR & T\\
    \hline
    ASE\_01 & 21.44 & -28.31 & 1.08 & 8.30 & -1.49 & -27.40\\
    ASE\_02 & 26.10 & -38.08 & 2.24 & 6.40 & -1.33 & -38.00\\
    ASE\_03 & 30.02 & -10.63 & 2.91 & 7.23 & -3.36 & -10.38\\
    ASE\_04 & 29.90 & -10.53 & 2.88 & 7.30 & -3.39 & -10.25\\
    ASE\_05 & 27.18 & -8.67 & 0.11 & 10.21 & -2.76 & -7.04\\
    ASE\_06 & 30.55 & -11.22 & 3.36 & 6.64 & -3.29 & -11.25\\
    ASE\_07 & 29.64 & -10.39 & 2.96 & 7.30 & -3.53 & -10.13\\
    ASE\_08 & 19.91 & -11.86 & -0.52 & 5.07 & -0.46 & -12.89\\
    ASE\_09 & 23.57 & -11.37 & 6.01 & 3.02 & -2.64 & -11.57\\
    \hline
    \end{tabular}
    \caption{Coefficients estimated for the SGWR model on standardized values}
    \label{tab:coeff_sgwr}
\end{table}

Then, let us look at the GWR calibration model outputs rather than the model parameters.

\subsection{Calibration results on non-collocated sensors}

\begin{figure}[H]
    \centering
    \includegraphics[width=\linewidth]{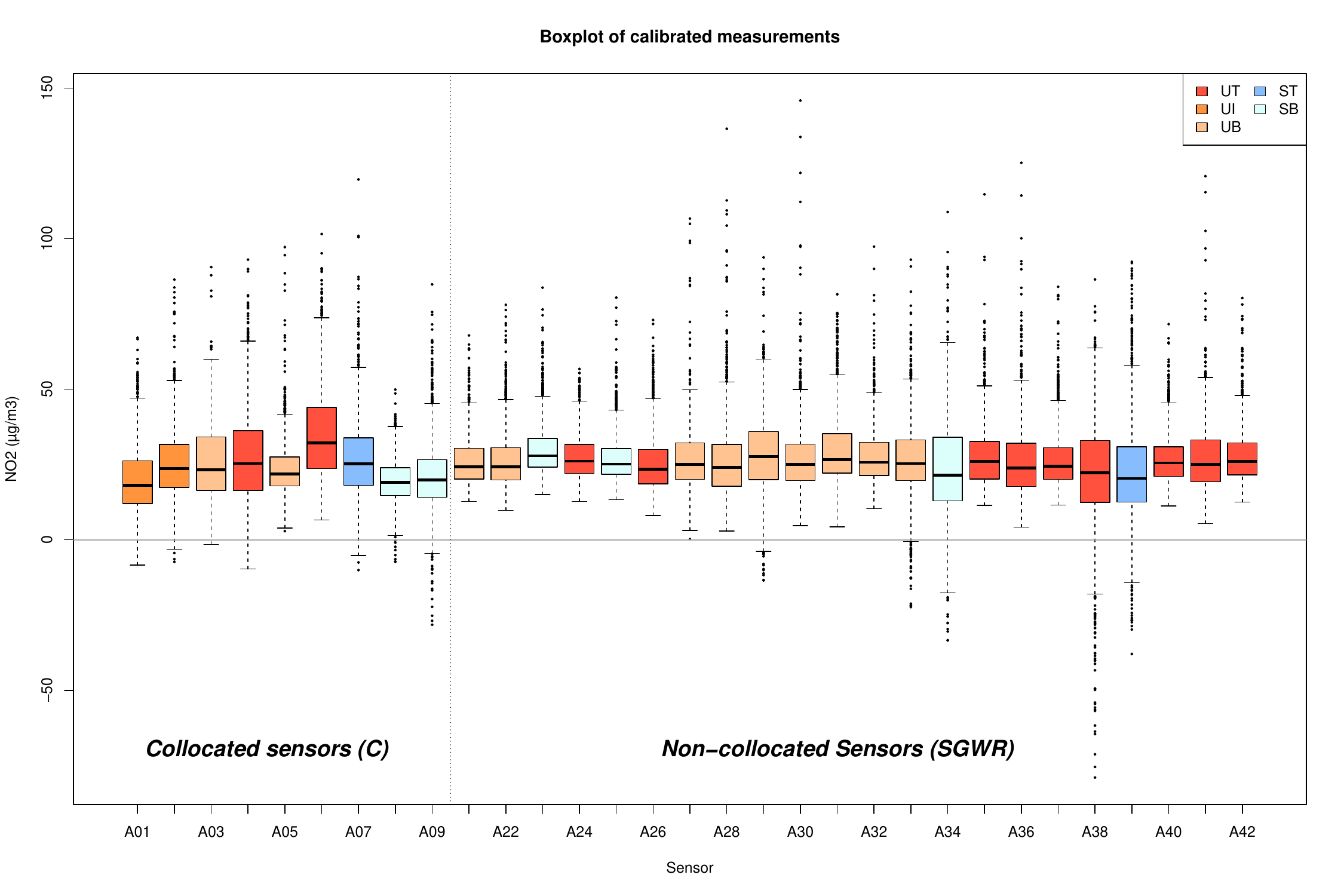}
    \caption{Boxplot of corrected measurements of sensors. The color of the boxplot relates to the location type: red (Urban Traffic), orange (Urban Industrial), salmon (Urban Background), blue (Suburban Traffic), light blue (Suburban Background). Non-collocated sensors are calibrated using SGWR, and for comparison, collocated sensors are calibrated using the best model possible, i.e. the collocated model.}
    \label{fig:boxplot_NC_calib}
\end{figure}

Figure \ref{fig:boxplot_NC_calib} shows the boxplots of the corrected measurements of non-collocated low-cost sensors over the test period (using the centering and reduction coefficients of the data from the learning period).

\noindent 
These results are extremely encouraging, in the sense that the corrected measurements give reasonable and realistic concentrations, with a limited number of negative values: only 5 boxplots out of 22. Out of approximately 1300 data points per sensor, the numbers of negative data are about 21 (ASE\_A29), 38 (ASE\_A33), 85 (ASE\_A34), 102 (ASE\_A38) and 107 (ASE\_A39). 
The last two sensors for which the results are the worst, are the two outlying low-cost sensors that are located around the industrial reference stations. 
These sensors are in a situation very different from these stations, which can explain the lower quality of the SGWR calibration model.
Moreover, when compared to the boxplots of the collocated calibration model, one can note that they exhibit similar shapes. Collocated models (which are considered the best linear calibration models) also produce a few negative measurements. Concentrations obtained using the SGWR model fall within the same range as those obtained from the collocated model.

\section{CONCLUSION AND DISCUSSION}

We focused on the use of the GWR framework 
to address in an unified way the calibration of low-cost sensors, from the learning and test sets choice to the final assessment using a spatial cross-validation scheme.
The calibration results for NO$_2$ together with some remarks about estimated GWR model, 
the spatial contents of estimated coefficients opens the assessment of the calibration of non-collocated low-cost sensors.

One of the key ingredient of the GWR method is the choice of the weight function and the most crucial parameter is the size of the window impacting directly the spatial resolution of the method. 
But, we could use weights to include more information about the problem.

\subsection{Time dependent weights}

A natural idea could be to include a time-dependent term in the weighting. 
This second function, detailed in Equation \eqref{eq:poids2}, will allow us to build a model called GTWR (see \cite{gwr:gtwr2015}), whose coefficients will depend not only on the location $s$ of the considered low-cost sensor but also on the hour $h$ of the day. The weight function is then :
\begin{equation}
 w_{j,t}(s,h) =
\underbrace{\exp\left(-\frac{1}{2}\frac{\lVert s - s_j\rVert^2_2}{B^2}\right)}_{\text{spatial dependency}} \times
\underbrace{\frac{1}{1+\left|h-h(t)\right|^3}}_{\text{time dependency}}
\label{eq:poids2}
\end{equation}
where $h(t)$ is the hour of the observation to weight, and other elements are the same as in Equation \eqref{eq:poids1}.
Note that in \cite{gwr:gtwr2015}, the spatial kernel is also an exponential, and our first experiments lead us to consider a polynomial kernel.

\noindent 
This dependency allows to put more weight on observations made in a similar time slot to the one at which we want to correct the measurement of the low-cost sensor. 

In our case, this would not be very interesting since the impact of the hour of the day on the performance seems to be small as shown by Figure \ref{fig:boxplot_CV_RMSE_h} which corresponds a similar idea with a very pricked function.

\subsection{Land-use weights}

\begin{figure}[H]
    \centering
    \includegraphics[width=0.48\linewidth]{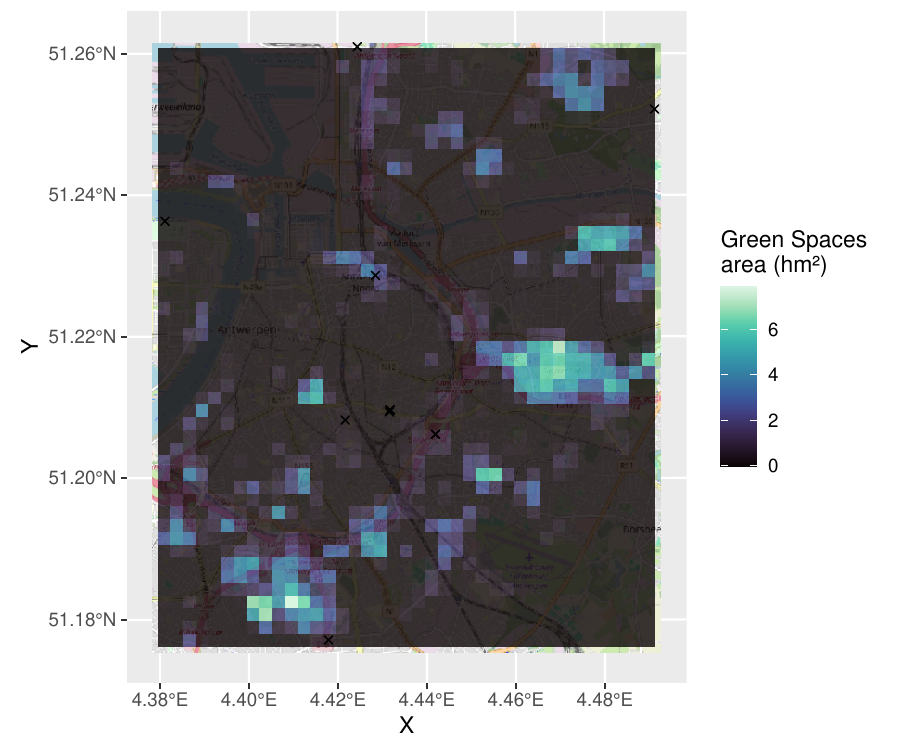} \hfill
    \includegraphics[width=0.48\linewidth]{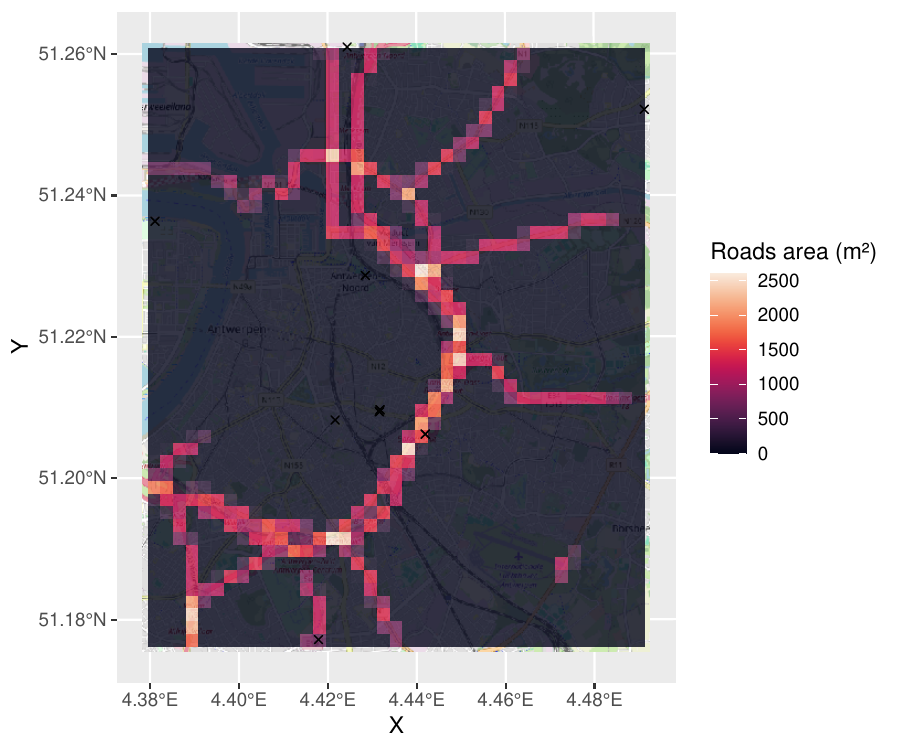}
    \caption{Spatial covariates used to characterize the sensor environment: green spaces (left) and roads (right) within a radius of 100 meters}
    \label{fig:landuse_var}
\end{figure}

Another idea could be include in weights some information related to land-use. 
This kind of spatial covariates designed to characterize the sites where the sensors are located could be used to model in a subtle way the typology of locations (urban, suburban, traffic, background, industry). For example, we exported the locations of roads and green spaces in the city of Antwerp from OpenStreetMap (see \cite{OpenStreetMap}) in order to calculate for each site the area represented by green spaces (resp. roads) within a radius of 100 meters. Processing was carried out using the package \texttt{sf} by \cite{sf}. 
These two cartographic variables are displayed in Figure \ref{fig:landuse_var}.

\noindent
It should be noted that cartographic variables are easy to integrate into the GWR formalism. 
But the GWR model allows the model to be adapted spatially via the spatial dependency of the parameters and, 
what is more, since these variables do not depend on time, the intercept will become unidentifiable.

\subsection{Transferable conclusions}

This work takes place in a scientific collaboration between Atmo Normandie, Paris-Saclay University and INSA Rouen Normandie. A previous study carried out by the authors in Rouen (see \cite{BobbiaStatistical2025}) address pointwise models. 
We mentioned at the end of the introduction that the main reason to use the Antwerp dataset instead of the ones currently available in Rouen is that the latter are limited. 
However, Antwerp has a very rich situation in terms of reference stations and low-cost sensors. 
Hence, it would be possible to extract two sub-networks analogous to Rouen's in order to draw transferable conclusions. Indeed, the two cities are similar in many respects, particularly with regard to meteorology, geography and economy: they are 350 km apart, and both are port cities. 
Furthermore, the sensors deployed in Antwerp are of the same AirSensEUR model as those in Rouen and measure the same pollutants. A way to quantify the expected quality is to evaluate the cross-validated error of a suitably chosen sub-network according to the scheme proposed in this paper.

\renewcommand{\refname}{REFERENCES}
\makeatletter
\renewcommand{\bibsection}{%
   \section*{\refname%
            \@mkboth{\MakeUppercase{\refname}}{\MakeUppercase{\refname}}%
   }
}
\makeatother
\renewcommand{\bibfont}{\small} 
\bibliographystyle{apalike}     
\bibliography{bibliographie.bib}   

\section*{APPENDIX. RENAMING SENSORS}
\label{app:nom-sensors}

To make it easier to work with this dataset and identify the various sensors, we have changed their names. The convention for this renaming is as follows:

\begin{itemize}
  \item Each sensor name starts by 'ASE\_A' followed by a number.
  \item Sensors numbered from 01 to 09 are collocated with a reference device during deployment.
  \item Sensors numbered 13, 14 et 16 are respectively collocated with sensors numbered 03, 04 et 06.
  \item Sensors numbered from 21 to 42 are non-collocated with reference devices during deployment.
\end{itemize}

Corresponding names can be found in Table \ref{tab:sensors_rename} below.
\begin{table}[H]
\centering
	\footnotesize
	\begin{tabular}{c|c}
		\textbf{Old name} & \textbf{New name} \\
		\hline
		4065DA & ASE\_A01 \\
		4065EA & ASE\_A02 \\
		4043B1 & ASE\_A03 \\
		4049A6 & ASE\_A04 \\
		4067BD & ASE\_A05 \\
		4043AE & ASE\_A06 \\
		4067B3 & ASE\_A07 \\
		40642B & ASE\_A08 \\
		4047D7 & ASE\_A09 \\
		40499C & ASE\_A13 \\
		4043A7 & ASE\_A14 \\
		40499F & ASE\_A16
	\end{tabular}
	\hfill
	\begin{tabular}{c|c}
		\textbf{Old name} & \textbf{New name} \\
		\hline
		406246 & ASE\_A21 \\
		4047CD & ASE\_A22 \\
		4065E0 & ASE\_A23 \\
		402B00 & ASE\_A24 \\
		4065D3 & ASE\_A25 \\
		4067BA & ASE\_A26 \\
		4065D0 & ASE\_A27 \\
		402723 & ASE\_A28 \\
		408168 & ASE\_A29 \\
		4047E7 & ASE\_A30 \\
		406424 & ASE\_A31 \\
		&
	\end{tabular}
	\hfill
	\begin{tabular}{c|c}
		\textbf{Old name} & \textbf{New name} \\
		\hline
		408165 & ASE\_A32 \\
		408175 & ASE\_A33 \\
		4047DD & ASE\_A34 \\
		408178 & ASE\_A35 \\
		4067B0 & ASE\_A36 \\
		4065DD & ASE\_A37 \\
		4065E3 & ASE\_A38 \\
		406249 & ASE\_A39 \\
		40623F & ASE\_A40 \\
		4047E0 & ASE\_A41 \\
		40641B & ASE\_A42 \\
		&
	\end{tabular}
	\caption{Sensors names, as they are described in the dataset (left columns) and in this article (right columns).}
	\label{tab:sensors_rename}
\end{table}
\end{document}